# Modulating Hierarchical Self-Assembly In Thermoresponsive Intrinsically Disordered Proteins Through High-Temperature Incubation Time

*Vaishali Sethi[1,2,3], Dana Cohen-Gerassi[2,3,4], Sagi Meir[1,2,3], Max Ney[5], Yulia Shmidov[5], Gil Koren[1,2,3], Lihi Adler-Abramovich[2,3,4], Ashutosh Chilkoti[5], and Roy Beck[*1,2,3]*

[1] School of Physics and Astronomy, Tel Aviv University, Tel Aviv 6997801, Israel.

[2] The Center for Physics and Chemistry of Living Systems, Tel Aviv University, Tel Aviv 6997801, Israel.

[3] The Center for Nanoscience and Nanotechnology, Tel Aviv University, Tel Aviv 6997801, Israel.

[4] Department of Oral Biology, The Goldschleger School of Dental Medicine, Sackler Faculty of Medicine, Tel Aviv University, Tel Aviv 6997801, Israel.

[5] Department of Biomedical Engineering, Duke University, Durham, NC 27708, USA

**Abstract**

The cornerstone of structural biology is the unique relationship between protein sequence and the 3D structure at equilibrium. Although intrinsically disordered proteins (IDPs) do not fold into a specific 3D structure, breaking this paradigm, some IDPs exhibit large-scale organization, such as liquid-liquid phase separation. In such cases, the structural plasticity has the potential to form numerous self-assembled structures out of thermal equilibrium. Here, we report that high-temperature incubation time is a defining parameter for micro and nanoscale self-assembly of resilin-like IDPs. Interestingly, high-resolution scanning electron microscopy micrographs reveal that an extended incubation time leads to the formation of micron-size rods and ellipsoids that depend on the amino acid sequence. More surprisingly, a prolonged incubation time also induces amino acid composition-dependent formation of short-range nanoscale order, such as periodic lamellar nanostructures. We, therefore, suggest that regulating the period of high-temperature incubation, in the one-phase regime, can serve as a unique method of controlling the hierarchical self-assembly mechanism of structurally disordered proteins.

**Introduction**

The organization of macromolecules into compartments allows cells tight regulation of physiological functions. In addition to classical membrane-bound organelles, eukaryotic cells also form dynamic, liquid-like membrane-less organelles (MLOs) to orchestrate biochemical reactions and cellular stress effectively.[1–3]

Intrinsically disordered proteins (IDPs) have significant stretches in their primary amino acid sequence-termed intrinsically disordered regions (IDRs) that lack a singular, stable 3D structure.[4-7] Such IDPs have attracted significant attention as hubs of protein-interaction networks that regulate transcription, translation, cell cycle[8–10], and modulate intracellular signaling.[11–13] In addition, liquid-liquid phase separation (LLPS) of IDPs has been shown to play an important role in the formation of MLOs,[14–16] and dysregulation of LLPS due to irreversible aggregation of IDPs can lead to a variety of pathological diseases.[17–19]

The LLPS of IDPs has also enabled diverse applications ranging from protein purification[20], drug delivery[21–23], self-assembly of artificial MLOs into nano-meso structures[24–26], to injectable biomaterials and hydrogels.[27–29] Importantly, beyond the sequence, other external stimuli, such as pH, protein concentration, and temperature, have also been shown to affect LLPS.[30]

Like synthetic polymers, IDPs can also exhibit thermoresponsive phase behavior, controlled by the spatial orientation of amino acid side chains and peptide-water interactions.[31–33] IDPs with an upper critical solution temperature (UCST) undergo LLPS below their transition temperature ($T_t$), also called the cloud point temperature. At high temperatures (T >$T_t$), the system is a homogenous solution, while at lower temperatures, it exists as a two-phase system consisting of an IDP-rich coacervate phase (immiscible in water) and a dilute phase of IDPs dissolved in the solvent. The UCST phase-separation is reversible, and above the $T_t$, the coacervates become soluble again.[20,34] In contrast, other IDP sequences with a lower critical solution temperature (LCST) exhibit phase separation above their cloud point temperature.[34] Examples of UCST and LCST IDP-based thermoresponsive polymers are resilin-like polypeptides (RLPs)[33] and elastin-like polypeptides (ELPs) [34], derived from the disordered sequences of the natural proteins resilin[35] and tropoelastin[36], respectively.

Designing new strategies to drive the self-assembly of resilin-like IDPs is important in light of their possible applications in tissue engineering, bioelectronics, bioimaging and biosensors.[37] Here, we investigate the thermoresponsive behavior and self-assembly of RLPs, a class of synthetic intrinsically disordered proteins (SynIDPs) derived from resilin.[35] We focus on their self-assembly into nano- and micro-scale structures modulated by changes in incubation time at temperatures above the LLPS transition temperature. In addition, we show that it is possible to fine-tune the resulting self-assembled structures by alterations in the sequence and the number of repeats of the RLPs.

**Results**

This study investigated the hierarchical self-assembly of SynIDPs derived from a 8 amino acid repeats (GRGDSPYS) inspired by the Drosophila Melanogaster Rec-1 resilin protein, which exhibits UCST phase behavior.[20] Here, we focused on three RLPs, the octapeptide parent sequence is known as Wild type (WT), while the notation for the RLPs are $[WT]_{XX}$, where XX represents the number of octapeptide repeats in the SynIDPs, for example, $[WT]_{40}$ and $[WT]_{80}$. Further, the terminology of mutant versions of the RLPs is according to the position and type of amino acids that is substituted for the amino acid into the WT octapeptide repeat. For example, the mutant $[V_7]_{40}$ has V substituted for the Y at the $7^{th}$ position in the octapeptide WT repeat. Specifically, we examined the effect of incubation time at 80° C ($t_{80}$), varying the number of octapeptide repeats and changing the sequence of the SynIDPs. All studied RLPs sequences are presented in Table 1.

Table 1. Characteristics of the A-IDP sequences

| Protein name | Sequence | No. of amino acids | Molecular weight (Da) |
|---|---|---|---|
| $[WT]_{20}$ | SKGP-(GRGDSPYS)$_{20}$-GY | 166 | 17,004 |
| $[WT]_{40}$ | SKGP-(GRGDSPYS)$_{40}$-GY | 326 | 33,400 |
| $[WT]_{60}$ | SKGP-(GRGDSPYS)$_{60}$-GY | 486 | 49,797 |
| $[WT]_{80}$ | SKGP-(GRGDSPYS)$_{80}$-GY | 646 | 66,193 |
| $[3Y_7V_7]_{40}$ | SKGP-[(GRGDSPYS)$_3$ GRGDSPVS]$_{10}$-GY | 326 | 32,760 |
| $[Y_7V_7]_{40}$ | SKGP-[GRGDSPYSGRGDSPVS]$_{20}$-GY | 326 | 32,119 |
| $[V_7]_{40}$ | SKGP-(GRGDSPVS)$_{40}$-GY | 326 | 30,839 |

**Microscale self-assembly**

Previous reports suggest that RLPs $[WT]_{40}$ and $[WT]_{80}$ show reversible LLPS below the cloud point, $T_t = 40°C$, when measured in 150 mM phosphate buffer of pH 7.4.[20] Both $[WT]_{40}$ and $[WT]_{80}$ RLPs are insoluble when prepared in buffer at 25°C. However, these samples do show LLPS after incubation at 80°C for $t_{80} = 1$ hr and cooling back to 25°C. In contrast, the mutant $[V_7]_{40}$ did not show such reversible UCST phase behavior and remained insoluble after 1 hr of incubation at 80°C.

We investigated the effect of incubation time on the microscale self-assembly of $[WT]_{40}$, $[WT]_{80}$, and $[V_7]_{40}$ by incubating 100 μM solutions of the RLPs in 150 mM phosphate buffer at 80°C for 1, 24, and 48 hr. After incubation, the solutions were cooled back to 25°C for 10 min. For convenience, we term a sample with XX repeats incubated at 80°C for $t_{80}$ hours and measured at 25°C as [sample]$_{XX}$-$t_{80}$.

High-resolution scanning electron microscopic (HRSEM) was used to characterize the morphology of the microstructures after different incubation times (Figs. 1-3). All the HRSEM were conducted after incubation at high temperature and cooling back to 25°C unless mentioned otherwise. Before hydration and self-assembly, the HRSEM images of $[WT]_{40}$ powder samples are amorphous (Fig. 1a). However, after hydration and 1 hr incubation at high temperature, images show that $[WT]_{40}$ are predominantly composed of 400 nm wide and ~600 nm long micro-ellipsoids with a few ~ 500 nm diameter rods that are several μm in length (Fig. 1b). A more prolonged incubation ($t_{80}$ = 24 hr), increased the number of rods in this system (Fig. 1c). Longer incubation ($t_{80}$ = 48 hr) produced an interconnected system of rods ~ 500 nm wide, but with length in microns (Fig. 1d). The micron-size ellipsoid particles are connected through inter-connecting nanofibers (arrows in Fig. 1b & Supplementary Fig. S1a). We speculate that these nanofibers grow radially and lead to the formation of micron size rods.

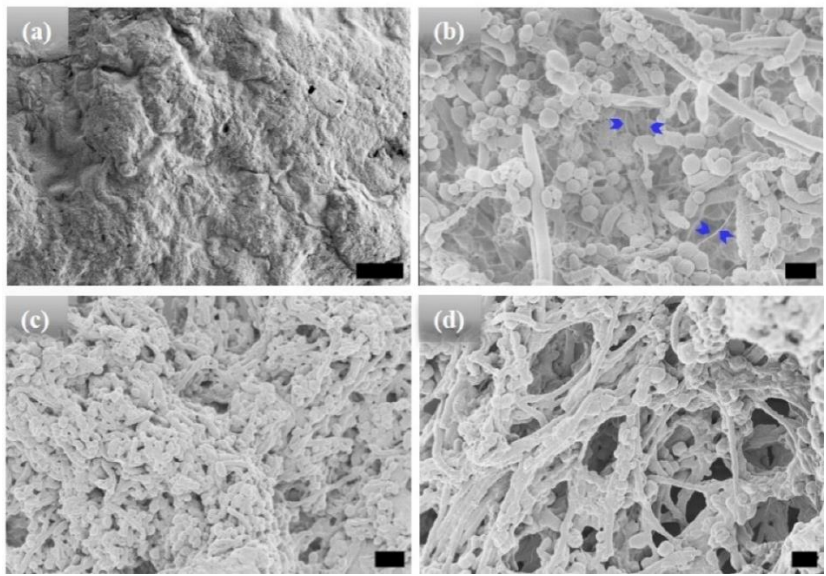

Figure 1. HRSEM micrographs of [WT]$_{40}$ RLP (a) amorphous powder samples. (b)-(d) [WT]$_{40}$ in phosphate buffer solutions after incubation at 80°C for 1, 24, and 48 hr, respectively. Incubation for 1 hr results in the formation of micro-ellipsoids of 500 nm width and ~1 μm in length and a few rods of ~ 500 nm in diameter and several μm in length (b). A longer incubation of 24 hr leads to an increase in the micro-rod population (c). A 48 hr incubation produced microstructures with interconnected rods (d). Arrows in (b) mark interconnecting nanofibers, which leads to self-assembly of micro-ellipsoids to form micro-rods. Scale bar: 1μm.

The [WT]$_{80}$ powder samples also exhibit an amorphous structure (Fig. 2a). However, imaging revealed that a hydrated [WT]$_{80}$ buffer solution only after incubation at 80°C for 1 hr and cooling back to 25°C, undergoes LLPS with the formation of condensates of several μm in length (Fig. 2b). Micro-rods and micro-ellipsoids were only seen after incubating the [WT]$_{80}$ peptide in buffer solution for 24 hr at 80°C (Fig. 2c). Similar to [WT]$_{40}$ RLP incubated for 48 hr, [WT]$_{80}$-24 hr sample also exhibits interconnecting nanofibers between the micro-ellipsoids which grow radially to form micron size rods (Supplementary Fig. S1b). Thus, in both [WT]$_{40}$ and [WT]$_{80}$ systems, 48 hr of incubation at 80°C induced the formation of long rods accompanied by a few micro-ellipsoids (Figs. 1d, 2d).

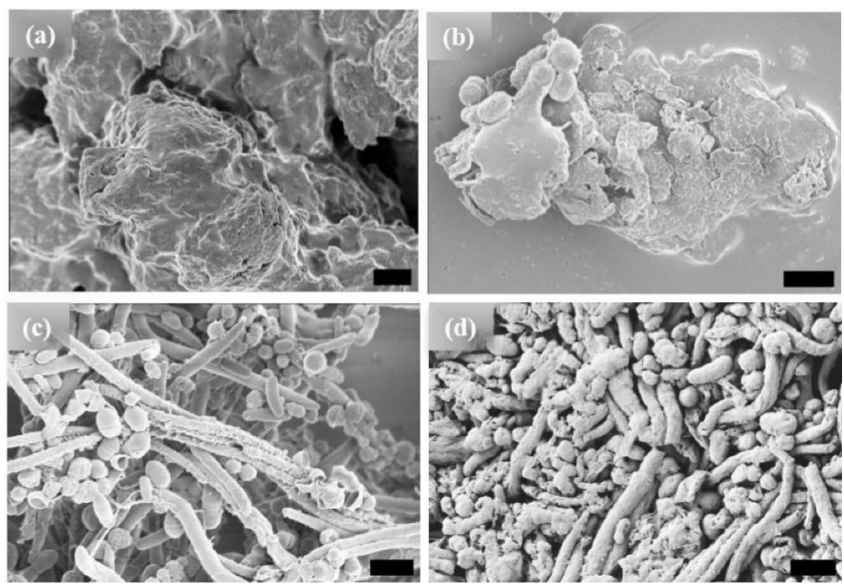

**Figure 2**. HRSEM images of [WT]$_{80}$ RLP (a) amorphous powder samples. Micrographs of RLP solutions in buffer after incubation at 80°C.(b) A 1hr incubation results in the formation of several μm long condensates, (c) 24 hr incubation produces a mixture of micro-rods and micro-ellipsoids, and (d) 48 hr incubation results in several μm long micro-rods accompanied by some micro-ellipsoids. Scale bar: 1μm.

Unlike the [WT] RLPs, the powder and 1 hr incubated samples (Fig. 3a and 3b respectively) of [V$_7$]$_{40}$ systems are insoluble in phosphate buffer and exhibits an amorphous structure. The sample becomes soluble in buffer and show LLPS after 24 and 48 hr of incubation at 80°C following cooling back to 25°C. After extended incubation time at high temperature (i.e., t$_{80}$ = 24 hr), the sample become turbid and exhibit the formation of small ellipsoidal particles (Fig. 3c). Upon incubating [V$_7$]$_{40}$ sample for t$_{80}$ = 48 hr ellipsoid particles of ~500nm in width and ~650 nm in length are observed by HRSEM with a few several μm long rods (Fig. 3d). Unlike [WT]$_{40}$, the primary morphology observed for [V$_7$]$_{40}$ system after 48 hr incubation is micro-ellipsoidal rather than elongated rods. No LLPS is observed in any of the three RLPs at 80°C, it starts only after the RLP solutions are cooled below their UCST. Further, we collected the HRSEM images (not shown) of all the three RLPs below UCST just after dissolution and they exhibit formation of amorphous structure similar to powder samples.

We further investigated the purity of the protein samples before incubation (BI) and after desired period of incubation through Matrix-assisted laser desorption/ionization time of flight mass spectrometry (MALDI-TOF-MS) studies. We witnessed the experimental m/z peaks for [WT]$_{40}$, [V$_7$]$_{40}$ and [WT]$_{80}$ before incubation samples at 33379, 30839 and 66193 respectively (Supplementary Fig. S2 (a), (c) and (e) respectively) corresponding to their molecular weight accompanied with a small peak for [WT]$_{40}$-BI and [V$_7$]$_{40}$-BI samples at m/z = 67135 and 61960 respectively. These small peaks corresponds to oligomer (dimer) peak which is common in MALDI-TOF-MS measurements and forms due to charge of the matrix.[38] The [WT]$_{40}$-48 hr, [V$_7$]$_{40}$-48 hr and [WT]$_{80}$-24 hr (Supplementary Fig. S2 (b), (d) and (f) respectively) samples exhibits single peak in their mass spectrum at m/z of 33379, 30839 and 66193 respectively, similar to what witnessed in before incubation samples.

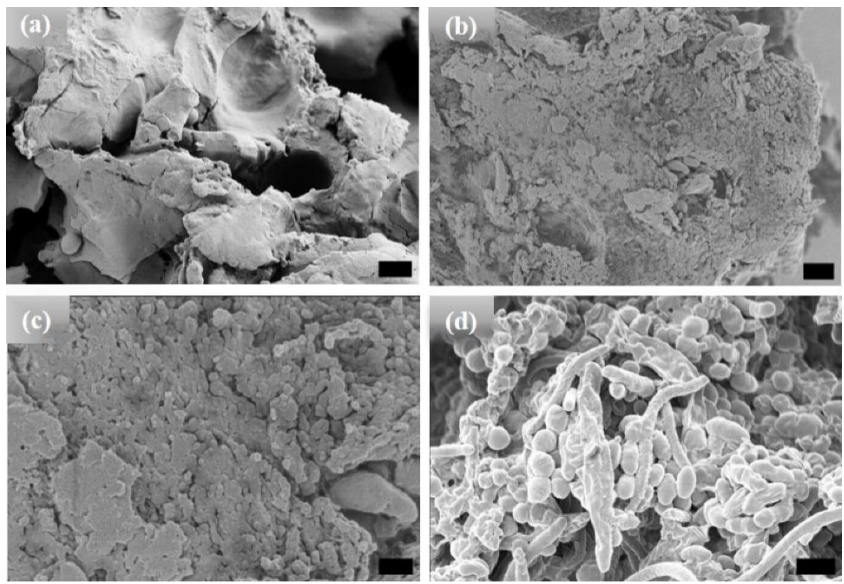

**Figure 3.** HRSEM images of [V$_7$]$_{40}$ RLP(a) powder samples, (b)-(d) [V$_7$]$_{40}$ RLP solutions after incubation at 80°C for 1, 24, and 48 h, respectively. The powder samples and solutions incubated for 1 and 24 hr have an amorphous structure, but micro-ellipsoids accompanied by few nanorods are observed after 48 hr incubation at 80°C. Scale bar: 1μm.

## **Nanoscopic Self-Assembly**

Motivated by the microscale self-assembly of SynIDPs and the correlation with incubation time, we used small- and wide-angle X-ray scattering (SAXS, WAXS) techniques to investigate the nanoscopic structures.[39,40] RLP samples and excess supernatant were sealed in polycarbonate capillaries and placed into a measurement heating stage. This allowed temperature dependent investigation of the samples with the desired high temperature incubation. For phase separated samples the X-ray scattering measurements is of the pellet positioned at the bottom of the capillary. The lack of correlation peaks in the scattering profiles of [WT]$_{40}$, [WT]$_{80}$, and [V$_7$]$_{40}$ solutions after 1 hr incubation (Fig. 4a) indicate that the structure is amorphous with no long- or short-range repeated order. Notably, there was no order in lyophilized powders of these A-IDPs (Supplementary Fig. S3a).

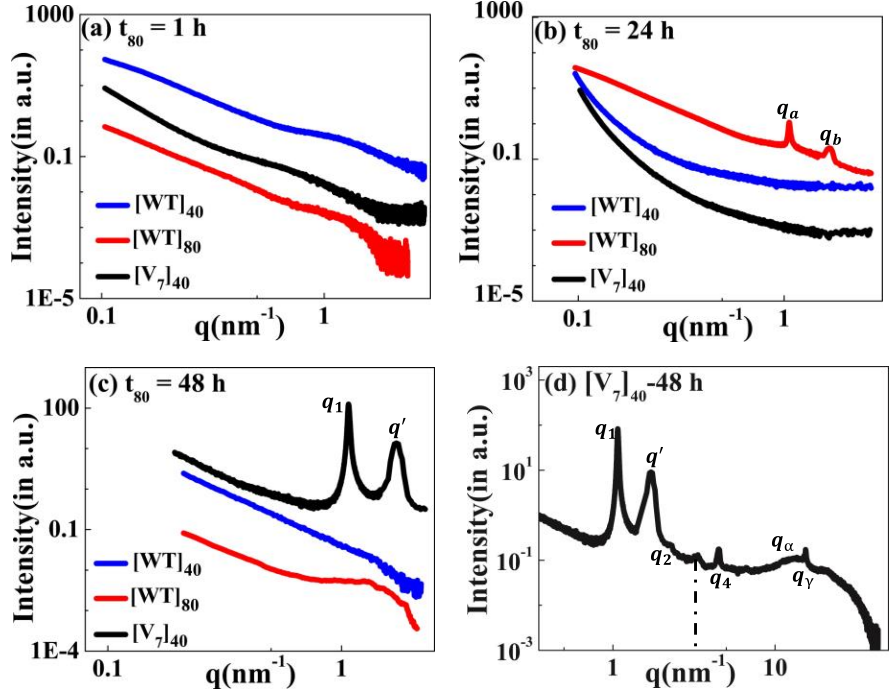

**Figure 4.** The scattering profiles of the $[WT]_{40}$, $[WT]_{80}$, and $[V_7]_{40}$ RLPs after incubation at 80°C ($t_{80}$) for the duration of (a) 1 hr, (b) 24 hr, and (c) 48 hr. There were no correlation peaks after 1 h but $[WT]_{80}$-24 hr and $[V_7]_{40}$-48 hr RLPs exhibit two intense correlation peaks $q_a$, $q_b$ and $q_1$, $q'$ respectively, indicative of nanoscopic long- or short-range ordering (d) Extended scattering profile of the $[V_7]_{40}$-48 h RLP exhibiting lamellar peaks ($q_1$, $q_2$, $q_4$), and a singular correlation peak ($q'$) with no additional reflections, and peaks $q_\alpha$, and $q_\gamma$. The peak at $q \sim 3.4$ nm$^{-1}$ (shown by a dotted line) corresponds to a residual diffraction peak of the kapton window originating from the experimental setup.[41]

To our surprise, incubation of $[WT]_{80}$ peptide solutions for $t_{80} = 24$ h resulted in the appearance of two sharp correlation peaks (i.e., $q_a$ and $q_b$) in the scattering profile (Fig. 4b). Such scattering peaks indicate the presence of repeated nanoscopic structures. Increasing the incubation time to 48 hr produced intense correlation peaks only in the $[V_7]_{40}$ -48 h system (Figs. 4c & 4d). Interestingly, the scattering profiles of $[3Y_7V_7]$ and $[Y_7V_7]$, where 25% and 50% of Y are replaced by V, respectively, do not exhibit correlation peaks after 48 h incubation (Supplementary Fig. S3b). Therefore, we can conclude that the both the sequence and incubation time ultimately dictate the nanoscopic long- or short-range self-assembly in RLPs considered for our studies. Specifically, the scattering profile of the $[V_7]_{40}$-48 hr system exhibits six correlation peaks ($q_1$, $q_2$, $q_4$, $q'$, $q_\alpha$, and $q_\gamma$, Fig. 4d) where the positions of peaks $q_1$, $q_2$ and $q_4$ are in the ratio $q_1:2q_1:4q_1$ and $q_1= 1.082$ nm$^{-1}$. This profile indicates the presence of an ordered lamellar morphology corresponding to an interplanar distance of 5.8 nm. In addition, there was a strong singular correlation peak, $q' = 1.646$ nm$^{-1}$, without any additional harmonics (Fig. 4d). The spacing of the correlation peaks at $q_\alpha=15.09$ nm$^{-1}$ and $q_\gamma=15.4$ nm$^{-1}$, corresponding to real space distances of $d_\alpha = 0.416$ nm and $d_\gamma = 0.406$ nm. Such sharp peaks at such length scale are indicative of ordering in the disordered amino acid side chain. The peaks $q_{\alpha_H}$ and $q_{\gamma_H}$ shifts towards lower $q$-values with increase in temperature during heating cycle from 25°C to 45°C (Supplementary Fig. S4 (a)). Furthermore, $q_{\alpha_H}$ and $q_{\gamma_H}$ peaks transform together to a hump from 50°C to 80°C (Supplementary Fig. S4 (b)). Upon cooling the system, we observed that hump exists from 70°C to 50°C (Supplementary Fig. S4 (c)) and the peaks $q_{\alpha_C}$ and $q_{\gamma_C}$ reappears at 45°C and shifts towards higher $q$-values down to 25°C (Supplementary Fig. S4 (d)). This indicates that during the heating cycle the correlation length for the peaks $q_{\alpha_H}$ and $q_{\gamma_H}$ increases and then becomes more disordered (hump), which happens due to breakdown of the of side chains order upon heating. Upon cooling again the partial ordering resume and the peak $q_{\alpha_C}$ and $q_{\gamma_C}$ reappeared. The peak positions of $q_{\alpha_H}$ and $q_{\gamma_H}$ are in close proximity to the characteristic spacing of *β*-sheets of 0.45-0.47nm[42], although they don't match it for clear identification as such.

We also performed circular dichroism (CD) studies for [V$_7$]$_{40}$ systems after various incubation times at high temperatures. After one hour of incubation the [V$_7$]$_{40}$-1hr RLP exhibits a peak at 197nm and rather flat spectrum through rest of wavelength considered. This is because [V$_7$]$_{40}$-1hr sample is insoluble in phosphate buffer. Thus, it's not possible to predict secondary structures from the CD spectra of [V$_7$]$_{40}$-1hr sample with certainty (Fig. 5a).

Following 48 hr incubation, the [V$_7$]$_{40}$ RLP did develop the characteristic minima at ~ 203 and 225 nm and a maxima at 190 nm similar to *β*-sheets reported for the Bombyx mori silk solution in hexafluoroisopropyl alcohol.[43] We acknowledge that for IDPs, the analysis of CD spectra is limited, since current models were derived based on folded proteins [44,45] thus we keep on suggesting that the nature of the peaks observed in the CD spectra of [V$_7$]$_{40}$-48 hr system is due to partial chain-ordering.

Acknowledging its limitation for IDPs, we used the BestSel algorithm to quantify the different types of secondary structures in the [V$_7$]$_{40}$ -48 hr RLP.[45] The quantification confirmed that the [V$_7$]$_{40}$ RLP after 48hrs of incubation exhibit a mixture of secondary structures including α-helix, parallel and antiparallel *β*-sheets, turns, and disordered coils (Supplementary Fig. S5), while the predominant structures are antiparallel *β*-sheets (34%, Supplementary Fig. S5).

Fourier transform infrared (FTIR) spectra of [V$_7$]$_{40}$ -48hr system exhibited a minima transmitted IR peak at 1637 cm$^{-1}$, and a weak minima at 1623 cm$^{-1}$ implies *β*-sheets (Fig. 5b).[46] The higher wavenumber band at 1732 cm$^{-1}$ corresponds to antiparallel *β*-sheets in the system in support with our CD data. Also, we didn't observe the amide bands in FTIR spectra of [V$_7$]$_{40}$ -1hr sample (Fig. 5b) because it is insoluble in phosphate buffer and thus are not deposited on KBr cards. Nonetheless, FTIR also have limitations in determining β-sheet structures correctly[46] and thus we can conclude that partial chain ordering in our system do appear and that it structure is likely to be similar to β-sheet formation.

In support of our experimental results, alphafold2[47] predicts completely disordered structure for $[WT]_{40}$ system (Supplementary Figs. S6 (a) which is infact is the characteristic of IDP. On the contrary, alphafold2 for $[V7]_{40}$ shows more than 90% probability of presence of β-sheets (Supplementary Fig. S6 (b)).

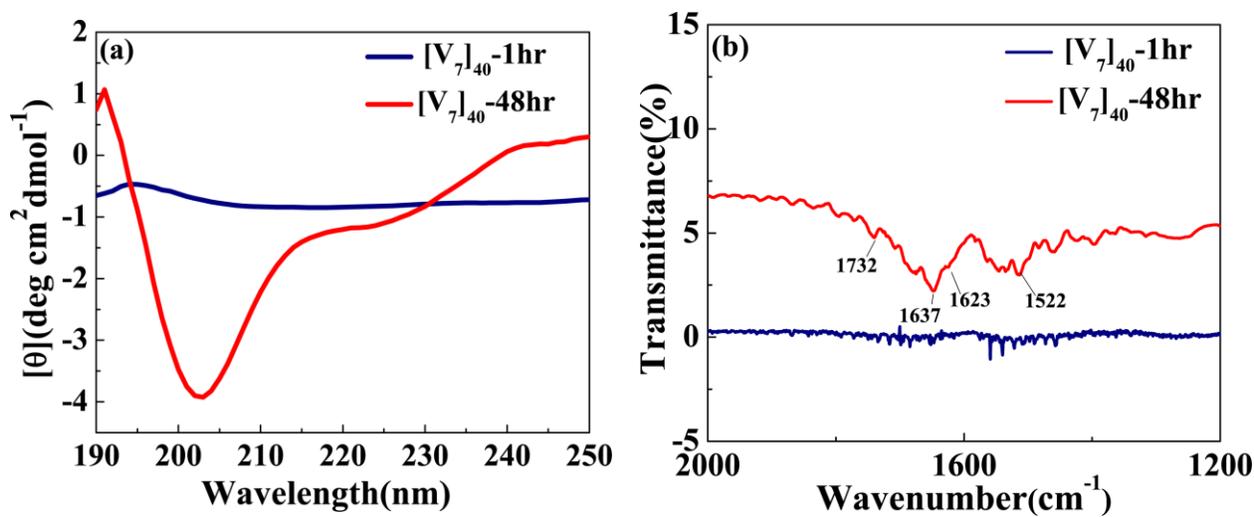

**Figure 5.** (a) CD spectra for $[V_7]_{40}$ RLP (a) After 1 and 48 hr incubation. The $[V_7]_{40}$-48 hr sample exhibits profile with minimas at the wavelengths of ~203 nm and ~225nm and a maxima at 190 nm characteristic to β-sheets. (b) FTIR spectra of $[V_7]_{40}$-1hr and $[V_7]_{40}$-48 hr sample. The presence of infrared bands at 1623 1637 and 1732 cm$^{-1}$ in $[V_7]_{40}$-48hr sample further indicates the possibility of the presence of antiparallel β-sheets.

Next, we used X-ray scattering to investigate the temperature-dependence phase behavior of the lamellar structure in the $[V_7]_{40}$-48 hr samples during the heating and cooling cycle. The experiment involves incubating at 80°C for $t_{80}$ hours and then cooling back to 25°C. The temperature ramped up to 80°C with a ramp rate of 1°C min$^{-1}$ during the heating cycle. After every 5°C increment, the system was held at the given temperature for 15 min for equilibration before collecting the scattering profiles (Supplementary Fig. S7). Scattering profiles were collected every 5°C during the cooling cycle from 80°C to 20°C (Supplementary Fig. S8).

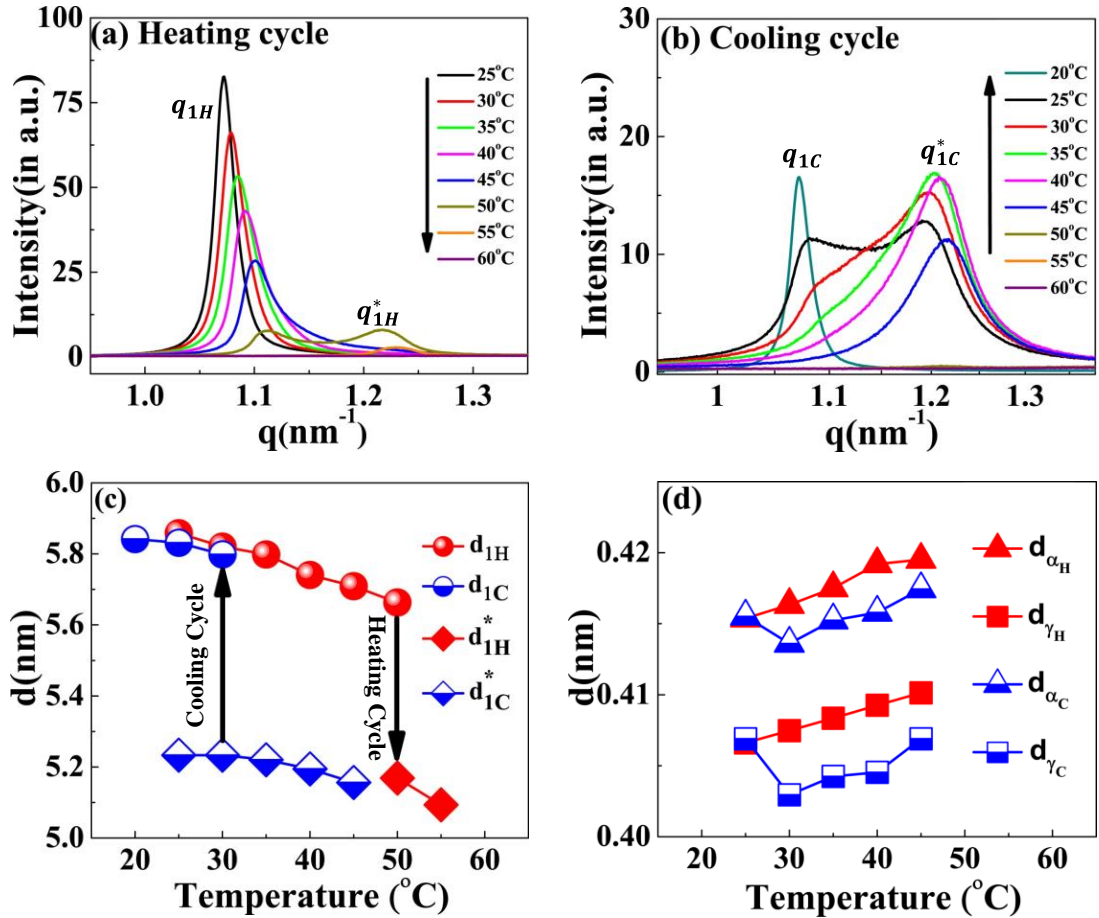

Figure 6. SAXS profiles of the $[V_7]_{40}$-48 h RLP at different temperatures during temperature cycle (a) heating and (b) cooling. With an increase in temperature, the principal lamellar peak ($q_{1_H}$) systematically moves toward higher q-values and splits at 50°C ($q_{1_H}^*$) and both correlation peaks disappear at 60°C. During the cooling cycle, a correlation peak reappears at 50°C as $q_{1_C}^*$, which denotes a thermal hysteresis of 15°C in the self-assembly. The peak $q_{1_C}^*$ spilts to a daughter peak starting as a small hump at 35°C and becoming prominent at 30°C & 25°C. Variation in the interplanar distance (d in nm) during the heating-cooling cycle corresponding to correlation peaks (c) $q_{1_H}, q_{1_H}^*$, $q_{1_C}$, and $q_{1_C}^*$ (d) $q_\alpha$ and $q_\gamma$ obtained by the relation d = $2\pi/q$. Arrows in (c) indicate the temperatures at which the system transit to the two-nanoscopic phase during the heating and cooling cycle.

The results indicate that while heating the $[V_7]_{40}$-48 hr sample from 25°C-45°C, the lamellar correlation peaks systematically shift towards higher *q*-values. However, at temperatures above 45°C, the higher-order lamellar harmonics ($q_{2_H}$, $q_{4_H}$) disappear. Upon further heating the system to 50°C, the higher harmonics disappears, while the principal lamellar peak ($q_{1_H}$), persists (Fig. 6a & Supplementary Fig. S7). Further heating the system to 50°C, causes $q_{1_H}$ to split into two peaks, where we designate the daughter peak as $q_{1_H}^*$. Surprisingly, on further heating peak $q_{1_H}$ disappears at 55°C, while $q_{1_H}^*$ persists. From 60°C to 80°C (the maximum temperature reached in our experiment), $q_{1_H}^*$ completely disappears (Fig. 6a). Above 60°C, the last ordered structure signature is the correlation peak of $q'_H$ that also remains in the one-phase regime (Fig. 6a & Supplementary Fig. S7). Further details of $q'_H$ will be discussed in the subsequent section.

There were no signs of lamellar correlation peaks when the temperature of the $[V_7]_{40}$-48 hr sample was decreased from 80 to 50°C during the cooling cycle (Fig. 6b & Supplementary Fig. S8). However, once the temperature reaches 45°C, we observe a single reflection, denoted as $q^*_{1_C}$, in the SAXS region (Fig. 6b & Supplementary Fig. S8).

Also, at 45°C, all the correlation peaks corresponding to a lamellar nanostructure reappear ($q^*_{1_C}$, $q_{2_C}$, $q_{4_C}$, where C in the subscript signifies the cooling cycle, see Supplementary Fig. S8). During further cooling, the lamellar peaks systematically shift to lower q-values (Fig. 6b & Supplementary Fig. S8). In addition, starting at 35°C, and seen very clearly at 25°C, the peak splits into two reflections $q^*_{1_C}$ (1.07 nm$^{-1}$) and $q_{1_C}$ (1.19 nm$^{-1}$), similar to those observed in the heat cycle at 50°C.

Furthermore, lowering the temperature to 20° C diminishes the peak at $q^*_{1_C}$ while $q_{1_C}$ peak persists (Fig. 6b). The lamellar interplanar distance (d = 2π/q) hysteresis loop during the heating and cooling cycle is presented in Fig. 6c. We could detect only a weak anti-correlation between the temperature and the interplanar distance but could observe the coexistence of two nanoscopic phases at 50° C and 30° C for the heating and cooling cycles, respectively. Noticeably, the $q_\alpha$ and $q_\gamma$ spacing i.e. $d_\alpha$ and $d_\gamma$ positively correlates with temperature between 25 to 45° C and disappears above this temperature (Fig. 6d).

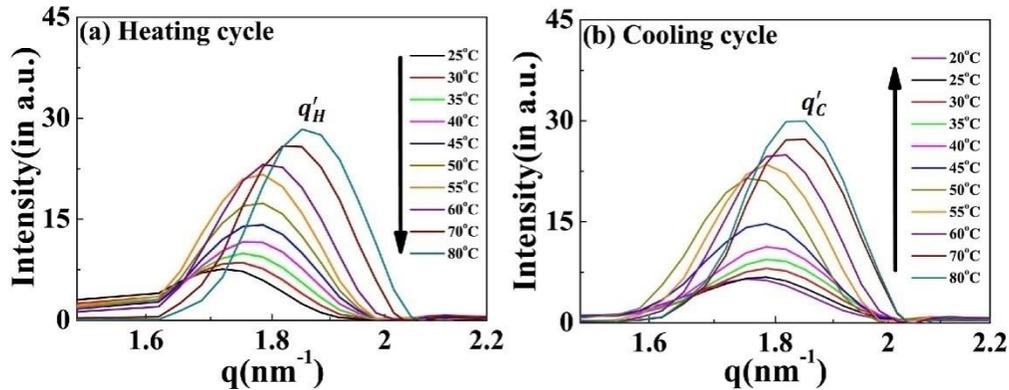

**Figure 7.** Scattering profiles of the $[V_7]_{40}$-48 h RLP exhibiting a variation in correlation peak $q'$ as a function of temperature. Profiles during the (a) heating cycle ($q'_H$) and (b) cooling cycle ($q'_C$). The $q'$ peak persists through all the temperature range examined, and the peak position is essentially independent of the heating and cooling cycle. The intensity of $q'$ peak increases with an increase in temperature for both cycles.

The correlation peak designated as $q'$ remains present throughout the complete heating-cooling cycle (Fig. 7a, b). Interestingly, the intensity of the $q'$ peak is positively correlated with the temperature during both cycles, with the peak position (and hence the inter-planer distance) independent of the cycle direction (Supplementary Fig. S9). Both these findings differentiate this from the other nanoscopic structures and suggest that a discrete equilibrated state is present in the system.

We also performed temperature-dependent CD studies for $[V_7]_{40}$-48 hr system with same experimental conditions as for SAXS measurements. Upon ramping the temperature from 20 to 80° C at a ramp rate of 5° C, we observed that the minimas at ~ 203 and 225 nm corresponding to partial ordering of side chain persist up to 60° C during heat cycle. These minimas almost vanish at 70 and 80° C (Supplementary Fig. S10 (a): for clarity we have shown results at interval of 10° C). During cool cycle the minimas and maxima reappeared at 60° C and persist up to 20° C (Supplementary Fig. S10 (b)).

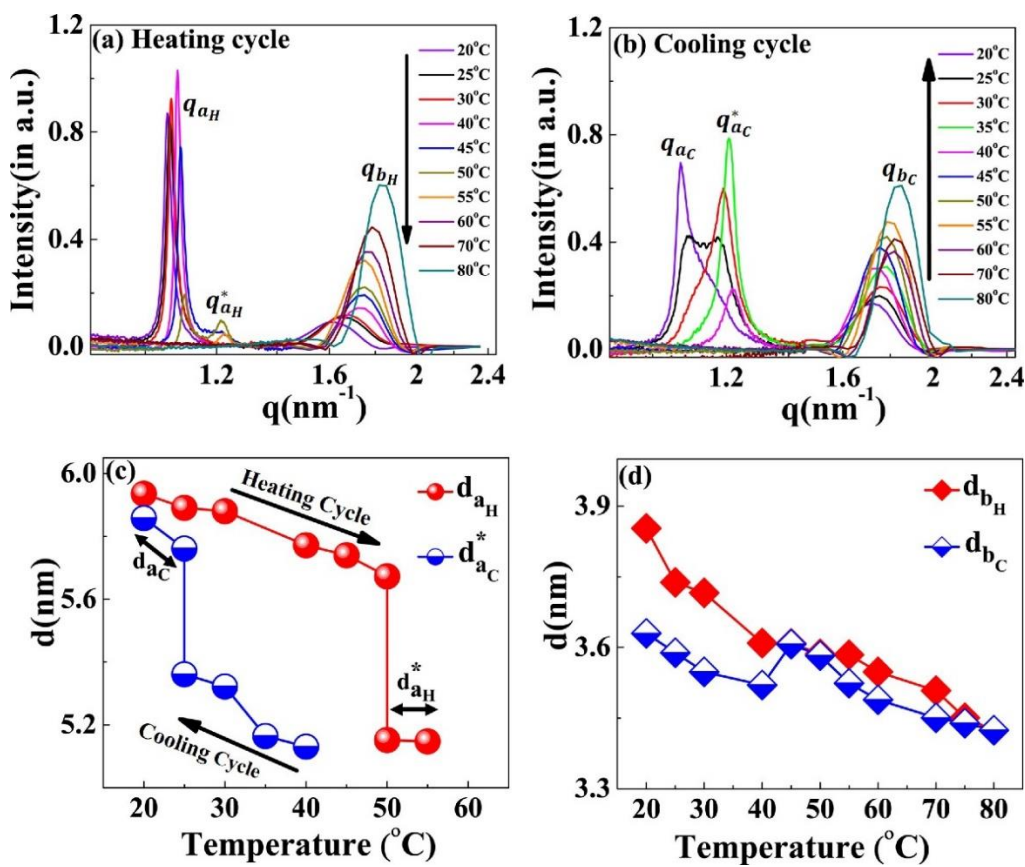

**Figure 8.** Scattering profiles of the $[WT]_{80}$-24 hr sample at different temperatures. During the (a) heating and (b) cooling cycle. The correlation peaks $q_a$ and $q_b$ indicates short-range nanoscopic organization. Upon ramping the temperature from 20°C- 45°C, the peak $q_{a_H}$ shifts towards higher q-values, splits into a daughter peak ($q^*_{a_H}$) at 50°C and both peaks disappear at 60°C. Upon cooling, the correlation peak reappears first at 40°C as $q^*_{a_C}$. This exhbits a thermal hysteresis of 20°C in the self-assembly. Below30°C, the $q^*_{a_C}$ diminishes and a new sharp correlation peak, $q_{a_C}$ appears at 20°C. Variations in the interplanar distance (d in nm) (c) $d_a$ (d) $d_b$ during the heating and cooling cycle.The system exhibits the presence of two-nanoscopic phases characterized by two-interplanar distances, $d_{a_H}$ and $d^*_{a_H}$ at 50°C and 55°C during the heating cycle and by $d_{a_C}$ and $d^*_{a_C}$ at 25°C during the cooling cycle. The correlation peak $q_b$ remains prominent throughout the heating-cooling cycle, with minimal changes in interplanar distance.

The scattering profiles of the [WT]$_n$ RLPs with a variable number of octapeptide repeats after 24 h incubation at 80°C differ from that of [V$_7$]$_{40}$ RLP. We could not detect any correlation peaks in the scattering profiles where n = 20, 40, and 60, (Supplementary Fig. S11), which indicates the absence of short-or-long range-order in these RLPs. However, the SAXS/WAXS profiles of the [WT]$_{80}$-24 h sample do contain two correlation peaks (Fig. 4b & Supplementary Fig. S12 (a) & (b)). These correlation peaks do not show corresponding higher-order reflections during both the heating and cooling cycle (Supplementary Fig. S12 (a) & (b)), thus preventing a description of the long-range symmetry. Noticeably the correlation peaks are rather sharp which suggests a homogeneous short-range correlation. Furthermore, we didn't observed correlation peaks in [WT]$_{80}$-48 hr sample (Fig. 4c) which might be due to disruption of the short-range correlation in this system upon increasing the incubation time from 24 to 48 hr.

During the heating cycle of the [WT]$_{80}$-24 hr sample, $q_{a_H}$ shifts systematically toward higher q-values and split into a daughter peak $q^*_{a_H}$ at 50°C (Fig. 8a) before both disappear at 55°C and 60°C respectively. In contrast, during the cooling cycle, the correlation peak ($q^*_{a_C}$) reappears at 40°C (Fig. 8b). Upon further cooling below 30°C, the correlation peak $q^*_{a_C}$ diminishes and a new sharp correlation peak, $q_{a_C}$= 1.07 nm$^{-1}$ appears and becomes more prominent as the system cools down to 20°C (Fig. 8b). Thus, we can conclude that the disappearance of $q^*_{a_H}$ at 60°C during the heating cycle and the appearance of $q_{a_C}$ at 40°C in the cooling cycle are interrelated.

Similar to the results seen with [V$_7$]$_{40}$, the [WT]$_{80}$-24 hr system exhibits thermal hysteresis of the nanoscopic structure (Fig. 8c). However, the exact short-range nanoscopic organization differs. While a previous report suggested that UCST SynIDPs like [WT]$_{80}$ show phase separation thermal hysteresis[20], our results demonstrate that nanoscale organization also exhibits such hysteretic behavior. In addition, the correlation peak $q_b$ remains prominent with little alteration and minimal hysteresis throughout the entire heating and cooling cycle, as shown by minimal changes in the interplanar distance (Fig. 8d).

## Discussion

The results presented here demonstrate modulation of the micro and nano self-assembly of resilin-like SynIDPs by periods of high-temperature incubation at 80°C and subsequent cooling. For the [WT]$_{40}$ and [WT]$_{80}$ RLPs, varying the incubation time correlates with a structural transition from micro-ellipsoids to rods structures. In contrast, the [V$_7$]$_{40}$ RLP, that has a V instead of Y amino acid at the 7$^{th}$ position of the GRGDSPYS repeat unit of WT RLP, requires a longer incubation time and only transitions from an amorphous coacervate to an ellipsoidal structure after 48 h at 80°C. Interestingly, upon cooling back to 25°C, prolonged incubation of both the [WT]$_{80}$, and [V$_7$]$_{40}$ RLPs at 80°C leads to the formation of hierarchical self-assembled nanostructures, an observation that was confirmed by the sharp X-ray scattering correlation peaks.

Based on their absorbance measurements, Quiroz et al.[34] reported that RLPs undergo loss of LLPS during multiple heating and cooling cycles around the UCST. They attributed such behavior to the formation of irreversible aggregates. A similar effect has also been observed for synthetic polymers exhibiting UCST phase behavior.[48] The loss of LLPS and the formation of irreversible pathological aggregates have also been observed in the low complexity disordered domain of RNA-binding protein FUS.[49] IDPs of Aβ, tau and α-synuclein proteins in aqueous solution also form such irreversible aggregates.[50] In line with these findings, we expect long incubations at 80°C to lead to a progressive build-up of irreversible aggregates in the [WT]$_{40}$ and [WT]$_{80}$ RLPs. Indeed, on the microscopic scale, we did observe various ellipsoidal aggregates in the HRSEM images after long incubation times at 80°C (Figs. 1c, d & 2c, d). Moreover, these ellipsoidal aggregates systematically self-assemble to form rods in the [WT]$_{40}$ and [WT]$_{80}$ RLPs.

The driving force for this self-assembly could be the abundance of H-bond's forming amino acids aspartic acid (D), serine (S) and tyrosine (Y) that can create metastable interchain interactions. Thus, we suggest that a long incubation time promotes additional mixing in the liquid state that can overcomes meta-stable states and accelerate the formation of self-assembled micro-scale structures via the hydrogen-bond's forming amino acids. These interactions could also explain the formation of interconnected nanofibers observed in these systems (Supplementary Fig S1).

Upon prolonged incubation at 80°C, the micro-ellipsoids aggregate throughout the nanofibers' connection to form rods or fiber-like structures. In contrast and supporting our hypothesis that intermolecular H-bonding is an important driving force for the hierarchical self-assembly in RLPs at higher temperatures is the lack of formation of interconnected fibers in [V$_7$]$_{40}$ RLP incubated for 48 h at 80°C. The [V$_7$]$_{40}$ RLP, has a non-H-bonding residue (V) instead of a H-bonding amino acid (Y). Further support is the reduction in the number of rods after prolonged 80°C incubation.

The incubation time also promotes order on the nanoscopic length scale in a sequence-dependent manner. The sequence of [V$_7$]$_{40}$ consists of a zwitterionic hydrophilic domain (i.e., GRGD), with positively charged arginine (R) and negatively charged aspartic acid (D). Each hydrophilic repeat is conjugated to a neutral-charged domain (i.e., SPVS) with a hydrophobic valine (V) residue. Thus, we presume that our sequence exhibits properties similar to amphiphilic polypeptides that possess alternating repeats of single hydrophilic and hydrophobic amino acids instead of alternating domains, each four amino acids long.[51,52] Alternating amphiphilic peptides are known to form highly stable $\beta$-sheet structures in aqueous solution due to the interaction between their hydrophobic faces,[51] while random polymers of the same composition lead to $\alpha$-helical structures.[53,54] Although, our CD and FTIR data (Figs. 5a & 5b respectively) points towards the presence of $\beta$-sheets in the $[V_7]_{40}$ RLP, our WAXS data do not point to typical $\beta$-sheets spacing. Hence, we suggest that the amphiphilic nature of our polypeptide may lead to partial side-chain ordering in these systems, which is most-likely has similarity to ordinary $\beta$-sheets.

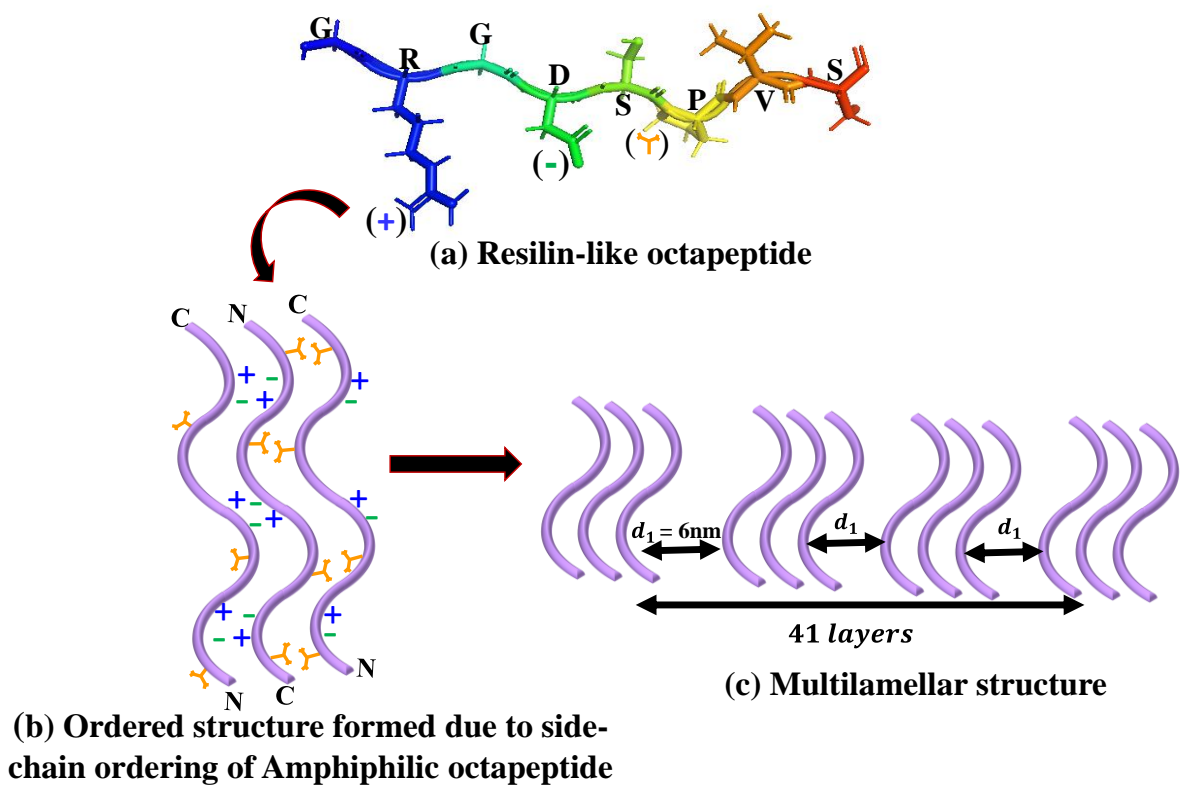

**Figure 9**. Schematic representation of self-assembly of hierarchical lamellar nanostructures in the $[V_7]_{40}$-48 hr RLP (a) PyMOL structure of a single RLP unit GRGDSPVS having hydrophobic and hydrophilic domain due to the presence of arginine (R) and aspartic acid (D), and valine (V) amino acids respectively. The (+) and (-) represent the charges on R and D and used to denote these amino acids in (b). At the same time, the orange Y-shape symbol describes the hydrophobic valine (b) Ordered structure formed due to partial ordering in side chains due to the alternating arrangement of the ionic (R and D) and hydrophobic (V) amino side chains on each strand of the amphiphilic octapeptide repeat. (c) Multilamellar nanostructures having interplanar distance ($d_i$) of 6 nm formed by lateral arrangement of ordered structure of octapeptide repeats. The domain size of multilamellar structure present in this system is 248 nm which corresponds to 41 layers of ordered structure.

Notably, the alternating amphiphilic sequence with valine alone is insufficient to induce partial side chain ordering, prolonged incubation (i.e., 48 hr) at a high temperature (80°C) and cooling back to 25°C is essential.[55] Thus, the major driving force for side-chain ordering is the self-interaction between the hydrophobic valine side chains on neighboring peptides, which minimizes unfavorable contact with water molecules. The self-complementary ion pair interactions between the alternate positively charged arginine (R) and negatively charged aspartic acid (D) motifs may also assist in stabilizing such interactions. Apparently, incubation up to 48 h is insufficient to induce such partial side-chain ordering in the $[WT]_{40}$ RLP, which shares identical ionic pairing.

In Fig. 9a we show schematic representation of resilin-like octapeptide unit GRGDSPVS having hydrophobic and hydrophilic domains due to the presence of arginine (R) and aspartic acid (D), and valine (V) amino acids amino acids, respectively. Due to the amphiphilic nature of our system, the valine side chains of two neighboring strands can interact with each other to form a hydrophobic face on one side away from the water molecules. On the other side, electrostatic interactions between complementary ionic amino acid pairs stabilize and leads to the formation of ordered structure formed due to partial ordering of these side chains (Fig. 9b).

The presence of sharp lamellar peaks in $[V_7]_{40}$-48 hr system suggests long-range order in the system. We are expecting the presence of multilamellar liquid crystalline structure as reported in triacylglycerol layers of lipid droplets found in adipose tissue.[56] We have calculated the domain size of multilamellar crystalline structure of $[V_7]_{40}$-48 hr system having a inter-planer distance ($d_1$) of 6 nm between corresponding ordered structure. The domain size is given by $\varepsilon = \frac{2\pi}{FWHM \cos\frac{\lambda q}{4\pi}}$ where FWHM represents the full-width at half-maximum of principle lamellar peak and represents the wavelength of X-ray in angstrom.[57] On the basis of this equation the $[V_7]_{40}$-48 hr system has $\varepsilon = 248$ nm with distance of 6nm between ordered structures, which corresponds to 41 layers in total (Fig. 9c).

Of note, we also observed that high temperature diminishes the self-assembled nanoscopic structures in the $[V_7]_{40}$ and $[WT]_{80}$ RLPs in a hysteretic manner. Previous reports suggest that IDPs exhibit temperature-dependent collapse, quantified as a decrease in their radius of gyration.[58–60] This temperature-induced collapse in IDPs implies the presence of temperature-dependent interactions such as hydrophobic effects[61,62] or changes in solvation-free energy.[30] Interestingly, Wuttke et al.[63] reported that the most hydrophilic and charged IDP sequences exhibit a significant temperature-induced collapse. They also suggested that along with the classical hydrophobic effects, the temperature-dependent solvation-free energies of the hydrophilic amino acids play a vital role in determining the collapse. Zerge et al.[64]

simulated the solvent-accessible surface area (SASA) and showed that this was decreased by hydrophobic amino acids up to a critical temperature and then subsequently increased. For negatively charged amino acids (Asp;D and Glu;E), the SASA decreases with increasing temperature, which avoids water contact, while the cationic amino acid arginine(R) exhibits the opposite tendency.[63,64] Therefore, the temperature-dependent self-assembly in our $[V_7]_{40}$ and $[WT]_{80}$ RLPs can be attributed to the collapse of the valine (V) or tyrosine (Y) domains due to the enhanced hydrophobic effect caused by an increase in temperature. Since the RLPs have R and D residues, which exhibit opposing SASA temperature dependencies, we might expect the combined temperature dependency to be more complex and require further investigation.

## Summary

We have demonstrated that it is possible to modulate the hierarchical self-assembly in RLPs by simply adjusting the time of incubation at high temperatures. We also witnessed a temperature-dependent hysteresis in the self-assembled nanostructure of these SynIDPs that is determined by their sequence, and suggest that the lamellar nanostructures formed in this system are associated to partial side - chain orderings. A variation in incubation time also induced structural transitions from amorphous to rod/ellipsoids during the LLPS of these SynIDPs. In conclusion, our results introduce a new approach to access new forms of self-assembled structures by the interplay between the incubation time and hysteresis of these SynIDPs. The micro-phase separated nanoscale lamellar or short-range ordered structure observed in our RLPs can find application in heterogeneous catalysis, drug delivery and scaffolds to guide tissue regeneration, similar to block copolypeptides.

## Materials and Methods

### Protein Purification

The RLPs were recombinantly expressed from plasmid-borne genes in E.Coli, and were purified as reported previously.[20] Briefly, *E. coli* (BL21) cells were grown overnight in Terrific Broth (TB) culture medium, followed by induction with 500 μM IPTG. The bacteria were pelleted at 3500 RCF, resuspended in PBS, then lysed by sonication and centrifuged at 4°C to separate the supernatant and pellet containing the RLP of interest. The pellets were suspended in an equal volume of 8M urea in 150 mM PBS buffer and heated for 10 min in a water bath maintained at 37°C, followed by centrifugation at 20,000 RCF for 20 min. The collected supernatants were dialyzed against milli-Q water at 4°C for 48 h with two changes of dialysis water. The suspension obtained after dialysis was centrifuged at 3500 RCF for 10 min at 4°C to pellet the RLP. The protein pellets were then lyophilized for three days to obtain dry RLP and stored at

-80°C as powder. The molecular weight and purity of each protein was confirmed by matrix-assisted laser desorption/ionization time of flight mass spectrometry (MALDI-TOF-MS).

**Sample preparation**

For small and wide-angle X-ray studies (SAXS and WAXS), buffers for background subtraction were prepared for each RLP by using slide-A-lyzer MINI dialysis devices (3.5K MWCO). A 100 μM solution of each RLP, prepared in 2mL phosphate buffer (150 mM, at pH 7.4), was incubated at 60°C for 15 min. The hot solution was transferred to the MINI dialysis tubes and dialyzed against 50 mL of phosphate buffer for 4-5 hr at 60°C on a shaker. Maintaining a temperature above the $T_t$ (i.e., 40°C) was necessary to prevent phase separation. We secured the mouth of the dialysis tube with a cellophane sheet to prevent evaporation of the buffer, and the tube was further closed with the cap. The dialyzed buffers were stored at 4°C and used for SAXS measurements as required.

For the RLP measurements, we used 100 μM dialyzed RLP solutions of $[WT]_{40}$, $[WT]_{80}$, and $[V_7]_{40}$ in triplicate. We heated the RLP solutions at 80°C on a heating bath for an incubation time of 1, 24, or 48 h. We defined the samples based on RLP name and incubation time. For example, samples of $[WT]_{40}$ RLP heated at 80°C for 1, 24, and 48 h are referred to as $[WT]_{40}$-1 hr, $[WT]_{40}$-24 hr, and $[WT]_{40}$-48 hr, respectively. Detailed material and methods are given in supporting information.

**Data Availability**

All data generated or analyzed during this study are included in this published article.

## Author Information

V.S. and R.B. designed the experiments and wrote the manuscript. V.S. prepared all the figures for the manuscript and conducted SAXS and CD measurements and analysis. S.M. assisted in the synchrotron SAXS's measurements. V.S., D.C.G., G.K. and L.A.A. conducted the HRSEM and FTIR measurements. M.N., Y.S. and A.C. purified the proteins and conducted Mass spectroscopic measurements. All the authors reviewed and refined the manuscript.

## Corresponding Author

*roy@tauex.tau.ac.il.

## Acknowledgments

The synchrotron SAXS/WAXS data were collected on a beamline I22 at Diamond Light Source, Oxford, United Kingdom. We would like to thank beamline scientists Dr. Nick Terrill and Dr. Andy Smith for their assistance in using the beamlines. This work was supported by the National Science Foundation under Grant No. 1715627, the United States-Israel Bi-national Science Foundation under Grant No. 2020787, and the Israel Science Foundation under Grants No. 1454/20 and 453/17, by the NIH through Grant No. R35GM127042 and the AFOSR though Grant No. FA9550-20-1-0241 (SUB0000436) to A.C., and the European Research Council (ERC) under the European Union's Horizon 2020 research and innovation programme (grant agreement no. 948102) to L.A.-A.. D.C.-G. acknowledges the Marian Gertner Institute for Medical Nano systems at Tel Aviv University. D.C.-G. gratefully acknowledges the support of the Colton Foundation. The authors acknowledge the Chaoul Center for Nanoscale Systems at Tel Aviv University and the ADAMA Center for Novel Delivery Systems in Crop Protections for the use of instruments and staff assistance. The authors thank Yacov Kantor, Cecilia Leal and Uri Raviv for fruitful discussions and support.

## Competing Interest

The authors declare no competing interests.